\documentclass[letterpaper,journal]{IEEEtran}
\usepackage{amsmath,amssymb,amsfonts}
\usepackage{algorithmic}
\usepackage{algorithm}
\usepackage{array}
\usepackage[caption=false,font=normalsize,labelfont=sf,textfont=sf]{subfig}
\usepackage{textcomp}
\usepackage{stfloats}
\usepackage{url}
\usepackage{verbatim}
\usepackage{graphicx}
\usepackage{cite}
\usepackage{orcidlink}
\usepackage{tikz}

\newcommand\submittedtext{%
  \footnotesize This work has been submitted to the IEEE for possible publication. Copyright may be transferred without notice, after which this version may no longer be accessible.}

\newcommand\submittednotice{%
\begin{tikzpicture}[remember picture,overlay]
\node[anchor=south,yshift=10pt] at (current page.south) {\fbox{\parbox{\dimexpr0.65\textwidth-\fboxsep-\fboxrule\relax}{\submittedtext}}};
\end{tikzpicture}%
}

\newcommand{\encB}[1]{\emph{enc}^{\mathrm{B64}}\!\left(#1\right)}
\newcommand{\decB}[1]{\emph{dec}^{\mathrm{B64}}\!\left(#1\right)}
\newcommand{\encP}[1]{\emph{enc}^{\mathrm{poly}}\!\left(#1\right)}
\newcommand{\decP}[1]{\emph{dec}^{\mathrm{poly}}\!\left(#1\right)}

\newcommand{\ser}[1]{\emph{ser}^{\mathrm{bin}}\!\left(#1\right)}
\newcommand{\des}[1]{\emph{des}^{\mathrm{bin}}\!\left(#1\right)}
\newcommand{\selMinLogN}[1]{\mathsf{selMinLogN}\!\left(#1\right)}
\newcommand{\selParams}[1]{\mathsf{selParams}\!\left(#1\right)}
\newcommand{\genRecords}[1]{\mathsf{genRecords}\!\left(#1\right)}
\newcommand{\feasible}[1]{\mathsf{feasible}\!\left(#1\right)}

\begin{document}

\title{Bringing Private Reads to Hyperledger Fabric via Private Information Retrieval}

\author{Artur Iasenovets~\orcidlink{0009-0008-0008-3746}, 
Fei Tang~\orcidlink{0000-0002-0048-9876}, 
Huihui Zhu~\orcidlink{0009-0006-5701-1837}, 
Ping Wang~\orcidlink{0009-0003-3832-2494}, 
and Lei Liu~\orcidlink{0009-0001-3851-4854}

\thanks{Received April 19, 2021; revised August 16, 2021. This work was supported in part by the National Key Research and Development Program of 
China under Grant 2021YFF0704102, in part by the Key Project of Science and Technology Research by Chongqing Education Commission under 
Grant KJZD-K202400610, and in part by the Chongqing Natural Science Foundation General Project under Grant CSTB2025NSCQ-GPX1263. 
(\emph{Corresponding author: Fei Tang.})

Artur Iasenovets and Fei Tang are with the School of Cybersecurity and Information Law, 
Chongqing University of Posts and Telecommunications, Chongqing 400065, China (email: L202420002@stu.cqupt.edu.cn, tangfei@cqupt.edu.cn).
Huihui Zhu, Ping Wang, and Lei Liu are with the School of Computer Science and Technology, Chongqing University of Posts and Telecommunications, 
Chongqing 400065, China (email: zhhcqupt@163.com, D220201031@stu.cqupt.edu.cn, d240201029@stu.cqupt.edu.cn).
}}

\maketitle

\submittednotice

\begin{abstract}
  Permissioned blockchains ensure integrity and auditability of shared data but expose 
  query parameters to peers during read operations, creating privacy risks for organizations querying sensitive records. 
  This paper proposes a Private Information Retrieval (PIR) mechanism to enable \emph{private reads} from Hyperledger Fabric’s 
  world state, allowing endorsing peers to process encrypted queries without learning which record is accessed. We implement and benchmark a PIR-enabled 
  chaincode that performs ciphertext–plaintext (\emph{ct×pt}) homomorphic multiplication directly within \emph{evaluate} 
  transactions, preserving Fabric’s endorsement and audit semantics. The prototype achieves an average end-to-end latency 
  of 113~ms and a peer-side execution time below 42~ms, with approximately 2~MB of peer network traffic per private read in 
  development mode—reducible by half under in-process deployment. Storage profiling across three channel configurations shows 
  near-linear growth: block size increases from 77 kilobytes to 294 kilobytes and world-state from 112 kilobytes to 
  332 kilobytes as the ring dimension scales from 8,192 to 32,768 coefficients. Parameter analysis further indicates that 
  ring size and record length jointly constrain packing capacity, supporting up to 512~records of 64~bytes each 
  under the largest configuration. These results confirm the practicality of PIR-based \emph{private reads} in Fabric 
  for smaller, sensitive datasets and highlight future directions to optimize performance and scalability.
\end{abstract}

\section{INTRODUCTION}\label{sec:introduction}

\subsection{MOTIVATION AND PROBLEM STATEMENT}
\IEEEPARstart{B}{lockchain} \cite{nakamoto_bitcoin_2008} is a peer-to-peer network protocol that uses cryptographic primitives and 
consensus mechanism to create and maintain a distributed ledger of transactions, such as Ethereum \cite{wood_ethereum_2014} and 
Hyperledger Fabric \cite{androulaki_hyperledger_2018}.

Hyperledger Fabric (HLF or Fabric) is widely adopted in various domains, one of them being Cyber Threat Intelligence (CTI) sharing
\cite{dunnett_trusted_2022, huff_distributed_2021, allouche_trade_2021, gong_blocis_2020}, where multiple organizations 
exchange threat indicators and wish to keep their queries confidential not to be assosiated with specific threats or incidents.
Blockchain guarantees, however, primarily cover data integrity and auditability, not query privacy.
This issue was recently highlighted by the Ethereum's privacy special interest group in their
``PSE Roadmap: 2025 and Beyond''~\cite{ethereum_pse_2025}, where they also 
emphasized the need for \emph{private reads} next to private writes and private voting.

In Hyperledger Fabric~\cite{androulaki_hyperledger_2018}, the separation of \emph{evaluate} and \emph{submit} 
makes the read-privacy gap explicit: an \emph{evaluate} call is a read-only proposal sent to endorsing peers, 
which execute the chaincode and return results without committing to the ledger \cite{fabric_docs_a_nodate}.
Crucially, these peers still observe all function arguments and read-sets, creating a privacy risk if queries are sensitive.

This motivates us to target \emph{read} privacy issue in Fabric, defined as the ability to 
hide which record is being queried from endorsing peers during \emph{evaluate} type calls. 
Given problem description, an obvious approach is to construct and use Private Information Retrieval (PIR)~\cite{chor_private_1998} 
protocol, which would enable client to retrieve an item from a world state database without revealing to peer which item was requested. 
However, integrating PIR into Fabric's architecture introduces non-trivial challenges related to execution model compatibility,
communication-computation trade-offs, and parameter tuning.

\subsection{CHOOSING AN APPROPRIATE PIR SCHEME}
PIR protocols can broadly be categorized into two families: \emph{Information-Theoretic} (IT-PIR) and \emph{Computational} (CPIR) schemes. Later,
each familly can be further divided into: \emph{non-preprocessing} schemes that perform all computation online, 
and \emph{preprocessing-based} schemes that split the protocol into offline and online phases.

\textbf{Information-Theoretic.} IT-PIR schemes~\cite{chor_private_1998,beimel_breaking_2002,goldberg_improving_2007,devet_optimally_2012,demmler_raid_pir_2014} 
provide unconditional privacy by distributing the database across \textbf{multiple non-colluding servers}, a client then queries subsets of 
servers such that no single server learns the selection index. 
Though IT-PIR schemes offer strong privacy guarantees and low communication overhead, they require multiple non-colluding servers,
which is often impractical in consortium blockchains where all peers are typically honest-but-curious and collusion is a realistic threat.

\textbf{PIR without Preprocessing.}
Older~\cite{cachin_computationally_1999,gentry_single-database_2005,brakerski_efficient_2011,dong_fast_2014,aguilar-melchor_xpir_2016} and 
more recent non-preprocessing single-server schemes such as SealPIR~\cite{sealpir_2018}, FastPIR~\cite{fastpir_2021}, OnionPIR~\cite{onionpir_2021}, 
and Spiral~\cite{spiral_2022} rely on cryptographic assumptions, such as homomorphic encryption (HE) to achieve privacy 
using \textbf{single untrusted server}. 
These constructions require only one round of interaction—an encrypted query sent to the server and an encrypted response returned—making them 
well-suited to Fabric’s \emph{evaluate} model, where minimal state is preserved between calls and each peer executes chaincode statelessly.  
While HE-based protocols are computationally heavier, they avoid persistent client-specific state and additional communication rounds, 
which are incompatible with Fabric’s endorsement flow and deterministic transaction semantics. 

\textbf{PIR with Preprocessing.}
Beimel et al.~\cite{beimel_reducing_2004} introduced the concept of PIR with preprocessing, where an \textbf{offline preprocessing is 
performed before the actual query} phase to reduce online communication and computation costs.
Recent advances such as SimplePIR~\cite{simplepir_2023}, Piano~\cite{piano_pir_2024}, 
HintlessPIR~\cite{hintless_pir_2024}, and YPIR~\cite{ypir_2024} achieve sublinear or near-constant online time by introducing an offline phase 
where the client or server precomputes query-independent data, usually called ``hints''.
Other doubly-efficient constructions~\cite{depir_ring_lwe_2023, okada_towards_depir_2025, corrigan-gibbs_subpir_2020} 
shift most computation to preprocessing, trading online efficiency for large offline communication or storage.
While appealing for cloud-hosted servers, these techniques are ill-suited for permissioned blockchains for two reasons:
(i)~if the hints are stored client-side, each participant must predownload large portions of the world state to generate them, 
defeating Fabric’s lightweight-client model; and
(ii)~if hints are stored peer-side, the chaincode or world state would need to maintain per-client data, 
violating Fabric’s stateless transaction design.  
Moreover, these schemes introduce significant storage blowup: for instance, SimplePIR's~\cite{simplepir_2023} 
client must download a 121 MB hint for a 1 GB database, and Piano's~\cite{piano_pir_2024} client preprocessing requires full-database 
downloads in its initialization stage. Such assumptions are incompatible with blockchain environments, where deterministic, replayable, 
and stateless transaction execution is critical.

Although less computationally efficient in theory and limited in database capacity due to polynomial packing constraints, we argue that 
\textbf{HE-based PIR currently is the most viable option} for enabling \emph{private reads} in Fabric-like permissioned ledgers, for the following reasons:
(i) HE-based PIR requires no changes to Fabric’s consensus, core architecture, or endorsement policies, 
and maintains the stateless transaction model without per-client state on peers.
(ii) It avoids the need for multiple non-colluding servers (or endorsing peers), which is difficult to guarantee in a permissioned blockchain setting.
(iii) It avoids offline preprocessing phases, which are incompatible with Fabric’s on-demand \emph{evaluate} calls, 
and prevents storage blowup on clients or peers.
(iv) It supports single-round query-response interactions that naturally align with Fabric's \emph{evaluate} calls.
(v) It leverages existing homomorphic encryption libraries and, with appropriate parameter choices, achieves practical performance.

We further argue that the limitations of HE-based PIR, particularly its database capacity constraints due to polynomial packing,
are an acceptable trade-off for enabling \emph{private reads} in permissioned ledgers handling smaller amounts of more
sensitive records in multi-organization settings.

\subsection{RELATED WORK}
We now review prior efforts to integrate PIR with blockchain and distributed systems to enhance query privacy. 
Detailed comparison Table~\ref{tab:comparison-related} is available in Section~\ref{sec:discussion}.

Xiao et al.~\cite{xiao_cloak_2024} propose Cloak, a privacy-preserving blockchain query scheme that uses distributed point functions (DPF) 
and noise-based sub-requests to hide retrieval information, achieving communication costs of 0.2-0.5 KB and query latency of 0.4-0.5 ms 
for databases with 4-64 records of 16-32 bytes each.

Mazmudar et al.~\cite{mazmudar_peer2pir_2025} propose Peer2PIR, a privacy solution for IPFS using a combination of PAILLIERPIR and RLWEPIR schemes 
to hide query content across peer routing, provider advertisements, and content retrieval functionalities.
For relevant private content retrieval function, they achieve communication costs of 10-15~MB and latencies of 0.5-15~s for 
databases of 1,000 to 1,000,000 records with 256~KB content blocks using the Spiral~\cite{spiral_2022} protocol.

Kaihua et al.~\cite{Kaihua2019applying} design an SPV protocol for lightweight Bitcoin clients using a hybrid PIR scheme (combining IT-PIR and C-PIR)
to enhance query privacy over Bloom-filter-based methods. Their system uses temporally-partitioned databases (Address, Merkle Tree, and Transaction DBs) 
to optimize performance. For a single transaction verification in the weekly database, it achieves a communication cost of 666~KB and 
latency of 2.84~s for databases with 7,688-512,460 records of 62-876 bytes each.

Kumar et al. introduced a series of HLF frameworks that integrate Private Information Retrieval (PIR) across different domains, 
namely BRON~\cite{kumar_bron_2024} and DEBPIR\cite{kumar_debpir_2025}. Although posed as enabling PIR functionality,
their evaluations primarily focus on the system's write throughput for general transactions rather than the specific performance of PIR queries.

\textbf{Targeted gap.}
Permissioned ledgers like Hyperledger Fabric ensure integrity and auditability of shared data but leak function arguments and read-sets
to endorsing peers during \emph{evaluate} calls, creating privacy risks in multi-organisation settings.
While prior works have explored PIR integration in various blockchain and distributed systems contexts,
none have specifically addressed the challenge of enabling \emph{private reads} in Hyperledger Fabric without modifying its core architecture.
To fill this gap, we explore how to enable \emph{private reads} from Fabric’s world state by integrating a HE-based PIR mechanism directly into chaincode, 
and we evaluate its practicality through detailed benchmarks and parameter studies.

\subsection{KEY OBJECTIVE AND CONTRIBUTIONS}
In this work, we demonstrate that HE-based PIR can be implemented natively in chaincode to enable \emph{private reads} in 
Hyperledger Fabric under \emph{evaluate} calls and achieve practical performance under certain parametrization choices.

Hence, the main contributions of this work are:
\begin{enumerate}
    \item \textbf{Enabling Private Reads in Hyperledger Fabric.} 
    We introduce the first end-to-end design, implementation and evaluation of a private information retrieval (PIR) mechanism, 
    leveraging homomorphic encryption, natively integrated into Hyperledger Fabric chaincode. This enables clients to perform \emph{private read} 
    operations on the world state through \emph{evaluate} transactions, ensuring the queried key remains confidential from endorsing peers. 
    Our approach requires no modifications to Fabric’s core, demonstrating a practical, \emph{plug-in privacy layer} for state queries.
    \item \textbf{Polynomial database construction.}
    We formalize the mapping between Fabric’s key–value world state and a homomorphically encodable polynomial database, 
    deriving feasibility constraints that link record length, ring dimension, and slot allocation. 
    These relationships ensure correctness of homomorphic retrieval and guide practical parameter selection for real-world datasets.
    \item \textbf{Multi-channel approach.} 
    To overcome database size limits imposed by polynomial packing constraints, we propose a multi-channel architecture 
    where each channel hosts a specific HE parameter set and corresponding polynomial database instance, 
    enabling clients to select appropriate channels based on their query needs.
    \item \textbf{Comprehensive evaluation.} 
    We implement a prototype using the Lattigo~\cite{noauthor_lattigo_2024} library and benchmark cryptographic operations, 
    blockchain interactions, and end-to-end query latency under various configurations. Our results show that single-query latencies 
    remain practical for typical Fabric deployments.
    \item \textbf{Open-Source Release.} To encourage reproducibility, we release the full implementation of our Fabric 
    chaincode, client logic and benchmarks at the repository: \url{https://github.com/iasenovets/2_2_HLF_CPIR}.
\end{enumerate}

\subsection{ORGANIZATION}

The remainder of this paper is organized as follows: 
Section~\ref{sec:preliminaries} reviews Fabric privacy features, database limits, Homomorphic Encryption-based PIR, and notation.
Section~\ref{sec:proposed_system} describes our system and threat models, polynomial database construction, feasibility constraints, 
multi-channel architecture, workflow and algorithms.
Section~\ref{sec:evaluation} shows experimental setup, cryptographic and blockchain benchmarks, and overall system performance.
Section~\ref{sec:discussion} discusses limitations and future directions, and Section~\ref{sec:conclusion} concludes the paper.

\section{PRELIMINARIES}\label{sec:preliminaries}

\subsection{FABRIC NATIVE PRIVACY FEATURES}
We hereby acknowledge several native solutions for privacy that Hyperledger Fabric provides and explain why they do not fully address the 
query privacy problem.

\textbf{Separate Channels.} Fabric's multi-channel architecture~\cite{fabric_docs_a_nodate} isolates ledgers across subgroups of organizations, 
limiting which participants observe which data. 
However, channel separation controls \emph{who} sees a ledger, query intent remains visible to all endorsers of a channel.

\textbf{Private Data Collections (PDC).} PDCs~\cite{fabric_docs_a_nodate} restrict which organizations store and access private key–value pairs. 
The shared ledger records only hashes, while members of the collection hold plaintext. 
PDCs provide access control but still expose function arguments to endorsers inside the collection, leaving query patterns observable.

\textbf{Fabric Private Chaincode (FPC).} FPC~\cite{fabric_docs_a_nodate, brandenburger_blockchain_2018} executes chaincode within Intel SGX enclaves. 
Arguments and state are protected even from peer operators, but this requires Trusted Execution Environments (TEEs) and attestation, 
introducing additional hardware and trust assumptions.

\subsection{FABRIC STORAGE LIMITS}
We here address the question of how large a database can be stored in Fabric world state and ledger history, as this impacts the 
practical limits of our PIR construction.

\textbf{World state (LevelDB/CouchDB).}
Fabric imposes no hard limit on the number of key–value entries in world state; capacity depends on available disk space and peer I/O throughput.
In our implementation, each channel maintains a few small artifacts in world state that amount up to a 350~KB under typical parameters (see Section~\ref{sec:evaluation}), 
LevelDB is sufficient and preferable for performance and simplicity, while CouchDB~\cite{fabric_docs_a_nodate} remains an option to extend from 8~MB to 4~GB if needed.

\textbf{Ledger history (blockchain log).}
According to the documentation~\cite{fabric_docs_a_nodate}, block size is constrained by the ordering service configuration. 
By default, the Fabric orderer limits the serialized payload to \emph{AbsoluteMaxBytes~=~10~MB} (recommended under 49~MB given the gRPC ceiling of 100~MB), 
and typically aggregates up to \emph{MaxMessageCount~=~500} transactions per block or \emph{PreferredMaxBytes~=~2~MB}.
In our system, these limits affect only \emph{submit} transactions such as \texttt{InitLedger} or record updates. 
\emph{Evaluate} transactions (including \texttt{PIRQuery}) are read-only and do not generate blocks, thus unaffected by ordering or batching constraints.

\textbf{Implication.}
The effective capacity of a channel is governed primarily by cryptographic feasibility defined in Section~\ref{sec:proposed_system},
and the size of a single world-state value (i.e., \emph{$m_{DB}$}), rather than by Fabric’s block or database limits. 
For very large objects, an optional extension is to store them off-chain 
in IPFS~\cite{benet_ipfs_2014} while persisting only their content identifiers (CIDs) in world state, 
keeping the polynomial \emph{$m_{DB}$} as the structured component used for private retrieval.

\subsection{HE-BASED PIR}
\textbf{Homomorphic Encryption.}
Lattice-based homomorphic encryption schemes, such as Brakerski/ Fan-Vercauteren (BFV), Brakerski-Gentry-Vaikuntanathan 
(BGV), CKKS and TFHE~\cite{cryptoeprint:2012/144,cryptoeprint:2011/277,cheon_homomorphic_2017,cryptoeprint:2018/421} have emerged as 
foundational technologies for practical CPIR implementations.

We base our CPIR construction on the BGV scheme, a lattice-based homomorphic encryption based on 
the Ring Learning With Errors (RLWE) problem~\cite{peikert_lattice_2014}, which supports both addition and multiplication over ciphertexts.
The BGV scheme defines operations over two polynomial rings: 
a ciphertext ring $R_Q = \mathbb{Z}_Q[X]/(X^N+1)$ and a plaintext ring $R_T = \mathbb{Z}_T[X]/(X^N+1)$, 
both sharing the same dimension $N=2^{\log N}$. 
In our implementation, these rings are jointly specified by a single parameter literal 
$(\log N, \log Q_i, \log P_i, T)$ as provided by the Lattigo library \cite{noauthor_lattigo_2024} and paper~\cite{mouchet_multiparty_2020}.
The field $T$ determines $R_T$, while the modulus chain $(Q,P)$ and their bit-lengths $(\log Q_i, \log P_i)$ 
determine $R_Q$.

\textbf{CPIR Definition.}
Model the database as a vector $D=\{d_0,\dots,d_{n-1}\}$. To retrieve desired record $d_i$ without revealing $i$, 
the client forms a one-hot selection vector $\hat{v}_i$ defined as $\hat{v}_i= (0,\dots,0,1,0,\dots,0)$ where $1$ 
corresponds to the desired record at index $i$. 
He then encrypts it as $ct_q = \emph{Enc}_{pk}(\hat{v}_i)$ under a public key $pk$ and sends $ct_q$ to the server, which computes 
$ct_r = (ct_q \cdot D) = \emph{Enc}_{pk}(d_i)$, returning $ct_r$ to the client for decryption $d_i = \emph{Dec}_{sk}(ct_r)$ 
using his secret key $sk$, thus obtaining the desired record $d_i$ without revealing $i$ to the server, or peer in our case.

\textbf{CPIR Instantiation:}
We instantiate Computational PIR as a tuple of probabilistic polynomial-time algorithms (\emph{KeyGen}, 
\emph{Enc}, \emph{Eval}, \emph{Dec}):
\begin{itemize}
    \item $\emph{KeyGen}(\lambda) \rightarrow (pk, sk)$: On input the security parameter $\lambda$, output a public key $pk$ and a secret key $sk$.
    \item $\emph{Enc}_{pk}(\hat{v}_i) \rightarrow ct_q$: Given a windowed selection vector $\hat{v}_i \in \{0,1\}^N$, 
    encode it into the plaintext ring $R_T$ and encrypt to a query ciphertext $ct_q$ under $pk$.
    \item $\emph{Eval}(ct_q, \emph{$m_{DB}$}) \rightarrow ct_r$: Given $ct_q$ and the plaintext polynomial database $\emph{$m_{DB}$} \in R_T$, 
    homomorphically evaluate the product to obtain an encrypted response $ct_r \in R_Q$.
    \item $\emph{Dec}_{sk}(ct_r) \rightarrow d_i$: Using the secret key $sk$, decrypt the response ciphertext $ct_r$ to recover the desired record $d_i$.
\end{itemize}

\noindent \textbf{Protocol objective.} 
Correctness requires that for all $i \in [n]$,
\[
\emph{Dec}_{sk}\!\left(\emph{Eval}\!\left(\emph{Enc}_{pk}(\hat{v}_i), \emph{$m_{DB}$}\right)\right) = d_i.
\]

\noindent \textit{Remark (restricted operation set).} 
The full BGV scheme also provides $\texttt{EvalKeyGen}$ to generate relinearization and rotation keys,
supporting ciphertext–ciphertext multiplication ($ct \times ct$), automorphisms, and modulus switching.
Since the database \emph{$m_{DB}$} is stored in plaintext within world state,
our PIR construction only requires homomorphic ciphertext–plaintext multiplication ($ct \times pt$).
Further extension to ciphertext–ciphertext operations is possible but incurs additional overhead, as discussed in Section~\ref{sec:discussion}.

\subsection{NOTATION}
We summarize the main notation used throughout the paper in Table~\ref{tab:notation}.

\begin{table}[!htbp]
\caption{Notation}
\label{tab:notation}
\setlength{\tabcolsep}{6pt}
\renewcommand{\arraystretch}{1.1}
    \begin{tabular}{|l|l|}
    \hline
    \textbf{Symbol} & \textbf{Description} \\
    \hline
    $\lambda$ & Security parameter \\
    \hline
    $n$ & Database size; \\ & index domain $[n]=\{0,\dots,n-1\}$ \\
    \hline
    $D$ & Database records $\{d_0,\dots,d_{n-1}\}$ \\
    \hline
    \emph{$m_{DB}$} & Plaintext polynomial \\ & representation of $D$ \\
    \hline
    $\hat{v}_i$ & One-hot selector for index $i$ \\
    \hline
    ${v}_i$ & Windowed selector for index $i$ \\ & with $record_s$ contiguous ones \\
    \hline
    ${c}$ & Coefficient vector $(c_0,\dots,c_{N-1})$ \\
    \hline
    $J_i$ & Disjoint window for index $i$ \\ 
    \hline
    $\emph{Enc}_{pk}(\cdot)$, $\emph{Dec}_{sk}(\cdot)$ & Encrypt / Decrypt \\
    \hline
    $\emph{Eval}(\cdot)$ & Homomorphic evaluation \\ & (ct–pt multiply) \\
    \hline
    $pk,\, sk$ & Public / secret keys \\
    \hline
    $ct_q$ & Encrypted query $\emph{Enc}_{pk}(\hat{v}_i)$ \\
    \hline
    $ct_r$ & Encrypted response $\emph{Eval}(ct_q, \emph{$m_{DB}$})$ \\
    \hline
    $d_i$ & Decrypted record $\emph{Dec}_{sk}(ct_r)$  \\
    \hline
    $\emph{KeyGen}(\lambda)$ & Key generation $\!\to\!(pk,sk)$ \\
    \hline
    $N$ & Ring dimension \\
    \hline
    $\log N$ & Logarithm base 2 of ring \\ & dimension $N$ \\
    \hline
    $\log Q_i$ & Bit-lengths of primes \\ & forming modulus chain $Q$ \\
    \hline
    $\log P_i$ & Bit-lengths of special primes $P$ \\
    \hline
    $T$ & Plaintext modulus \\
    \hline
    $record_s$ & Slots allocated per record \\
    \hline
    $record_b$ & Base serialized size of a record in bytes \\
    \hline
    $record_{\mu,\log N}$ & Template-specific minimum \\
    \hline
    $\mathcal{S}$ & Allowed discrete slot sizes \\
    \hline
    $|\cdot|$, $\mathrm{size}(\cdot)$ & Length in elements / bytes \\
    \hline
    $\mathcal{C}(N,s,n)$ & Capacity predicate: $n \cdot s \leq N$ \\
    \hline
    $\mathcal{M}(\log N,s)$ & Template predicate: $s \geq record_{\mu,\log N}$ \\
    \hline
    $\mathcal{D}(s)$ & Discrete predicate: $s \in \mathcal{S}$ \\
    \hline
    $\mathcal{F}(\log N,s,n)$ & Feasibility tuple: $\mathcal{C} \wedge \mathcal{M} \wedge \mathcal{D}$ \\
    \hline
    $\mathcal{DO},\,\mathcal{DW},\,\mathcal{DR},\,\mathcal{GW}$ & Data Owner; Data Writer; \\ & Data Requester; Gateway \\
    \hline
    \emph{evaluate}, \emph{submit} & Fabric read / write transaction types \\
    \hline
    $\mathcal{L}$ & Leakage considered \\ & (ciphertext size, protocol timing) \\
    \hline
    \end{tabular}
\end{table}
    
\section{PROPOSED SYSTEM}\label{sec:proposed_system}
\subsection{SYSTEM MODEL}
We introduce a blockchain-based query privacy system designed for permissioned ledgers. 
The system enables clients to perform \emph{private reads} from the ledger while endorsing peers can evaluate 
read-only queries over encrypted inputs without learning which record was accessed.  
We achieve this by integrating a lattice-based CPIR scheme based on the
BGV~\cite{cryptoeprint:2011/277} homomorphic encryption scheme directly into Fabric chaincode.
This approach ensures that clients remain the sole holders of decryption keys, while peers perform only 
black-box computations, thereby enhancing overall privacy without requiring trusted hardware or protocol modifications.

\noindent Our system is composed of the following entities:

\begin{itemize}
    \item \textbf{Data Owner ($\mathcal{DO}$):} Endorsing peers that hold the current plaintext 
    polynomial \emph{$m_{DB}$} in world state and execute PIR during \emph{evaluate}. $\mathcal{DO}$ is honest-but-curious.
    \item \textbf{Data Writer ($\mathcal{DW}$):} A client organization that provisions or refreshes 
    the database. $\mathcal{DW}$ invokes \emph{submit} to initialize the ledger (e.g., set $n$ and template bounds). 
    Chaincode computes $record_s$, packs $D=\{d_0,\dots,d_{n-1}\}$, encodes it into \emph{$m_{DB}$}, and persists it.
    \item \textbf{Data Requester ($\mathcal{DR}$):} A client that privately retrieves a record. $\mathcal{DR}$ 
    runs $\emph{KeyGen}(\lambda)\!\to\!(pk,sk)$, forms $ct_q=\emph{Enc}_{pk}(v_i)$, calls \emph{evaluate} \texttt{PIRQuery}, and later decrypts $ct_r$.
    \item \textbf{Gateway ($\mathcal{GW}$):} The Fabric client/chaincode interface used by $\mathcal{DW}$ 
    and $\mathcal{DR}$ to invoke \texttt{InitLedger}, \texttt{GetMetadata}, and \texttt{PIRQuery}. It follows standard Fabric semantics; no extra trust is assumed.
\end{itemize}

\noindent\textit{Remark (world-state scope).} In Fabric, the ``ledger" comprises the blockchain log and the world state. 
Our CPIR operates on the world state: \emph{$m_{DB}$} encodes the latest key--value snapshot, not the historical transaction logs.
Both \emph{world state} and \emph{ledger} capacity concerns are addressed in Section~\ref{sec:preliminaries}.

\subsection{THREAT MODEL}

Our design follows the standard \emph{honest-but-curious} adversarial model. 
We explicitly consider the following assumptions and threats:

\begin{itemize}
    \item \textbf{Endorsing peers ($\mathcal{DO}$).} Execute chaincode correctly but may try to infer the queried index 
    from \emph{evaluate} inputs or logs. They see $ct_q$, metadata, and \emph{$m_{DB}$}.
    \item \textbf{Data Writer ($\mathcal{DW}$).} Issues initialization writes via \emph{submit}. $\mathcal{DW}$ is not 
    trusted with decryption keys and learns nothing about $\mathcal{DR}$’s queries. We assume $\mathcal{DW}$ follows the 
    write protocol but is not relied upon for privacy.
    \item \textbf{External observers.} May eavesdrop on client–peer traffic. Without $sk$, $ct_q$ and $ct_r$ reveal nothing under BGV assumptions.
\end{itemize}

\noindent \textbf{Security objective.} 
For any $i\in[n]$, neither $\mathcal{DO}$ nor external observers can distinguish which $d_i$ is requested from $ct_q$ and $ct_r$.
The only permissible leakage is ciphertext size and protocol timing, denoted collectively as $\mathcal{L}$.

\subsection{SYSTEM OVERVIEW}

\begin{figure}[!t]
    \centering
    \includegraphics[width=\columnwidth]{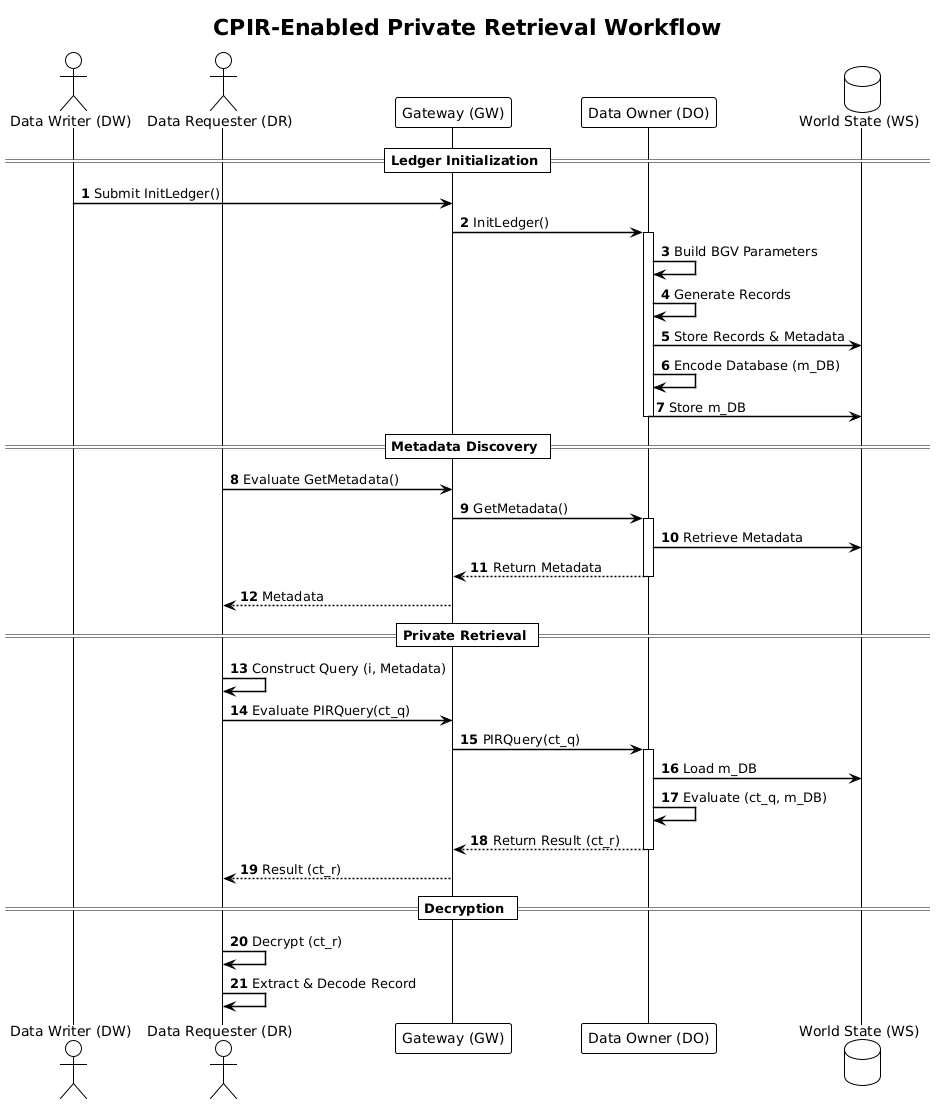}
    \caption{“Workflow. $\mathcal{DW}$ initializes the ledger via $\mathcal{GW}$, which triggers chaincode on endorsing 
    peers ($\mathcal{DO}$). $\mathcal{DO}$ executes the protocol and persists state in world state (\emph{$m_{DB}$}, metadata, 
    JSON records). $\mathcal{DR}$ later obtains metadata, submits $ct_q=Enc_{pk}(v_i)$, $\mathcal{DO}$ evaluates 
    $ct_r=Eval(ct_q,m{\mathrm{DB}})$ against world state, and $\mathcal{DR}$ decrypts to $d_i$.}
    \label{fig:system-overview}
\end{figure}

The proposed system integrates computational Private Information Retrieval (CPIR) directly into Hyperledger Fabric chaincode. 
Its purpose is to ensure that query indices remain hidden from endorsing peers while preserving Fabric’s endorsement and audit workflow. 
At a high level, the workflow consists of four stages, illustrated in Fig.~\ref{fig:system-overview}.

\begin{enumerate}
    \item \textbf{Ledger initialization.}
    $\mathcal{DW}$ invokes \texttt{InitLedger} via $\mathcal{GW}$ using \emph{submit}. Chaincode derives $record_s$ from 
    $record_b$, packs $D$ into ${c}=(c_0,\dots,c_{N-1})$, encodes \emph{$m_{DB}$}, and stores \emph{$m_{DB}$} and 
    metadata in world state held by $\mathcal{DO}$.
    \item \textbf{Metadata discovery.}
    $\mathcal{DR}$ calls \texttt{GetMetadata} via \emph{evaluate} to obtain $n$, $record_s$, and BGV parameters needed
    to form a valid query.
    \item \textbf{Private retrieval.}
    $\mathcal{DR}$ constructs $ct_q=\emph{Enc}_{pk}(v_i)$ and invokes \texttt{PIRQuery} via \emph{evaluate}. $\mathcal{DO}$ 
    computes $ct_r=\emph{Eval}(ct_q,\emph{$m_{DB}$})$ and returns it.
    \item \textbf{Decryption.}
    $\mathcal{DR}$ decrypts $ct_r$ to recover $d_i=\emph{Dec}_{sk}(ct_r)$.
\end{enumerate}

\subsection{POLYNOMIAL DATABASE CONSTRUCTION}

\begin{figure*}[!t]
    \centering
    \includegraphics[width=\textwidth]{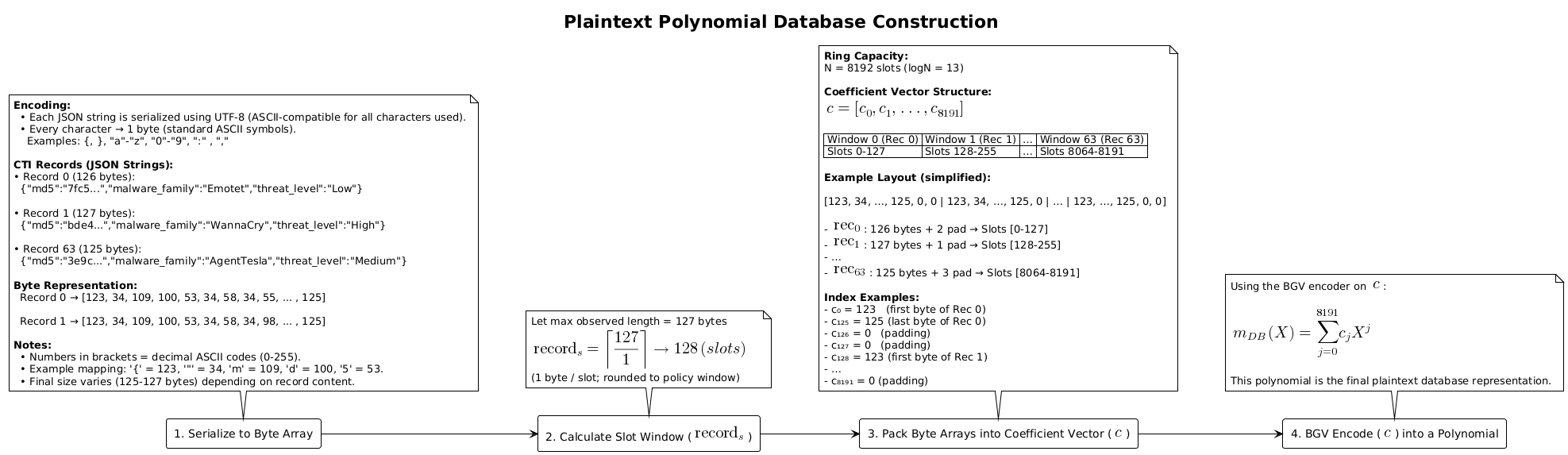}
    \caption{\emph{$m_{DB}$} construction from JSON to plaintext polynomial. 
    Each record is serialized to bytes, mapped into a fixed slot window $record_s$, and packed into a coefficient vector $c$. 
    The vector is then encoded as a BGV plaintext polynomial \emph{$m_{DB}$}, which is stored in the Fabric world state.}
    \label{fig:polynomial-construction}
\end{figure*}

To enable PIR queries over structured ledger data, we must embed records into a plaintext polynomial \emph{$m_{DB}$} suitable for BGV evaluation. 
Our prototype adopts a fixed-width packing strategy, illustrated in Fig.~\ref{fig:polynomial-construction}, which proceeds in four steps.

\textbf{Step 1: Serialize to Byte Array.}
Each structured record $d_i$ is serialized into a byte array using a deterministic JSON-to-bytes scheme.
Every character is represented by its ASCII code in $[0,255]$, which is also our plaintext modulus $T$. 
Although we set $T=65537$ as default in Section~\ref{sec:evaluation}, we use $T=256$ as an example here for clarity.

\textbf{Step 2: Calculate Slot Window.}
To achieve uniform packing, we determine a fixed slot window $record_s$ across all records, where $record_b$
is the maximum serialized record length in bytes and $bytesPerSlot$ is the number of bytes stored per slot.
We first calculate the basic slot requirement and then apply a discrete rounding policy:
\begin{equation}
\label{eq:record_s}
record_s = 8 \cdot \left\lceil \frac{record_b}{8 \cdot bytesPerSlot} \right\rceil,
\end{equation}
where $bytesPerSlot = \frac{\log_2(T)}{8}$ given plaintext modulus $T$. In our implementation, we set $T=256$, so $\log_2(256) = 8$ 
and thus $bytesPerSlot = 1$ byte. For example, if the largest record is $126$ bytes, then the basic requirement is $\lceil 126 / 1 \rceil = 126$ slots,
and after discrete rounding $record_s = 8 \cdot \lceil 126 / 8 \rceil = 8 \cdot 16 = 128$ slots.

\textbf{Step 3: Pack into Coefficient Vector.} 
Each serialized record $d_i$ is packed into its disjoint window $J_i = \{i \cdot record_s, \dots, (i+1) \cdot record_s - 1\}$ 
in the coefficient vector $c = (c_0, c_1, \dots, c_{N-1})$:
\begin{equation}
\label{eq:packing}
c[J_i[k]] = c[i \cdot record_s + k] = 
\begin{cases}
d_i[k] & \text{if } k < |d_i| \\
0 & \text{otherwise}
\end{cases}
\end{equation}
for $i \in [n]$ and $k \in [0, record_s-1]$, where $d_i[k]$ denotes the $k$-th byte of record $d_i$,
$J_i[k] = i \cdot record_s + k$ is the $k$-th slot index in window $J_i$ and
$c_j \in [0,T-1]$ are polynomial coefficients.

Padding zeros are added if a record is shorter than $record_s$. Thus each record $d_i$ occupies a contiguous slot interval that can 
be privately retrieved through PIR.

\textbf{Step 4: Encode into Polynomial.} 
Finally, the coefficient vector $c$ is encoded into a plaintext polynomial:
\begin{equation}
\label{eq:polynomial_db}
\emph{$m_{DB}$}(X) = \sum_{j=0}^{N-1} c_j X^j \in R_T,
\end{equation}
where $R_T = \mathbb{Z}_T[X]/(X^N+1)$. 
This polynomial serves as the database representation \emph{$m_{DB}$} in the PIR protocol, which endorsing peers use during query evaluation and
clients recover only the slots corresponding to their requested record.

\subsection{FEASIBILITY CONSTRAINTS}

Embedding records into the plaintext polynomial \emph{$m_{DB}$} is feasible only for parameter triples 
$(\log N, n, record_s)$ that satisfy \emph{all} of the following constraints. 

\textbf{Constraint 1: Ring capacity.}
The total number of occupied slots cannot exceed the ring size:
\begin{equation}
    \label{eq:ring_capacity}
    n \cdot record_s \leq N.
\end{equation}
This represents the fundamental mathematical limit imposed by the cryptographic parameters. 
For example, with $\log N = 13$ ($N = 8192$) and $record_s = 224$, 
at most $\lfloor 8192/224 \rfloor = 36$ records can be packed.

\textbf{Constraint 2: Template-specific minima.}
We anticipate that different application scenarios require different record templates composed of distinct field combinations. 
Each template $\mu$ imposes a minimum slot requirement $record_{\mu,\log N}$, determined by its mandatory fields:

\begin{equation}
    \label{eq:template_minima}
    record_{\mu,\log N} = B_\mu + \sum_{i \in \mathcal{H}_\mu} |F_i| + O_\mu,
\end{equation}
where $B_\mu$ is the base structure size for template $\mu$,
$\mathcal{H}_\mu \subseteq {F_1, F_2, \dots, F_L}$, denotes the set of fields included in template $\mu$,
$|F_i|$ is the byte length of field $F_i$,
$O_\mu$ captures serialization overhead and metadata specific to template $\mu$.

Plug in the CTI record templates we consider in this work in Eq.~\eqref{eq:template_minima}, we have:
\begin{itemize}
    \item \( record_{\mu,13} = \{ B_{\text{mini}}, F_{\text{MD5}} \} \), with \( B_{\text{mini}} \approx 81 \) bytes and 
    \( O = 15 \) bytes, so \( record_{\mu,13} \approx 81 + 32 + 15 = 128 \) bytes.
    \item \( record_{\mu,14} = \{ B_{\text{mid}}, F_{\text{MD5}}, F_{\text{SHA256-s}} \} \), with \( B_{\text{mid}} \approx 161 \) bytes and 
    \( O = 15 \) bytes, so \( record_{\mu,14} \approx 161 + 32 + 16 + 15 = 224 \) bytes.
    \item \( record_{\mu,15} = \{ B_{\text{rich}}, F_{\text{MD5}}, F_{\text{SHA256-l}} \} \), with \( B_{\text{rich}} \approx 145 \) bytes and 
    \( O = 15 \) bytes, so \( record_{\mu,15} \approx 145 + 32 + 64 + 15 = 256 \) bytes.
\end{itemize}

These minima ensure that for each template $\mu$, the slot allocation satisfies $record_s \geq record_{\mu,\log N}$, 
guaranteeing adequate space for all mandatory fields while maintaining the uniform packing strategy across records.

\textbf{Constraint 3: Discrete allocation policy.}
Operationally, we restrict the slot window $record_s$ to a discrete set for implementation simplicity:
\begin{equation}
    \label{eq:discrete_allocation}
    record_s \in \mathcal{S},
    \mathcal{S} = \{64, 128, 224, 256, 384, 512\} \ \text{bytes}.
\end{equation}
This means that only certain fixed slot sizes are allowed,
simplifying client query construction and chaincode logic.

\noindent\textbf{Constraint hierarchy and feasibility.}
We summarize the three constraints as predicates:
\begin{subequations}
    \begin{align}
    \mathcal{C}(N,s,n) & : n \cdot s \leq N \quad, \\
    \mathcal{M}(\log N,s) & : s \geq record_{\mu,\log N} \quad, \\
    \mathcal{D}(s) & : s \in \mathcal{S} \quad.
    \end{align}
\end{subequations}
where $\mathcal{C}$ captures ring capacity, $\mathcal{M}$ captures template minima,
and $\mathcal{D}$ captures discrete allocation, $s$ denotes $record_s$ here for brevity. The overall feasibility condition is:
\begin{equation}
    \label{eq:feasibility_condition}
    \mathcal{F} \iff
    \mathcal{C}(N,s,n)\ \wedge\ \mathcal{M}(\log N,s)\ \wedge\ \mathcal{D}(s)
\end{equation}

\noindent\textbf{Examples.} We provide three concrete examples below:

\noindent\emph{a) Polynomial degree} $\log N=13$ (Mini):
\begin{itemize}
\item $\mathcal{M}(13,s) \Rightarrow s \geq record_{\mu,13} \approx 128$ $\Rightarrow$ the smallest candidate is $s=128$.
\item $\mathcal{D}(s) \Rightarrow s \in \mathcal{S}$; $64 \notin \mathcal{S}$ (fails $\mathcal{M}$: $64 < 128$), $128 \in \mathcal{S}$.
\item $\mathcal{C}(8192,128,n) \Rightarrow n \leq \lfloor 8192/128 \rfloor = 64$.
\item $\mathcal{F}: (\log N=13,s \geq 128,n \leq 64)$.
\end{itemize}

\noindent\emph{b) Polynomial degree} $\log N=14$ (Mid):
\begin{itemize}
\item $\mathcal{M}(14,s) \Rightarrow s \geq record_{\mu,14} \approx 224$ $\Rightarrow$ smallest candidate is $224$.
\item $\mathcal{D}(s) \Rightarrow s \in {224,256,\dots} \subset \mathcal{S}$; $64 \notin \mathcal{S}$ (fails $\mathcal{M}$), $128 \notin \mathcal{S}$ (fails record generation).
\item $\mathcal{C}(16384,224,n) \Rightarrow n \leq \lfloor 16384/224 \rfloor = 73$.
\item $\mathcal{F}: (\log N=14,s \geq 224,n \leq 73)$.
\end{itemize}

\noindent\emph{c) Polynomial degree} $\log N=15$ (Rich):
\begin{itemize}
\item $\mathcal{M}(15,s) \Rightarrow s \geq record_{\mu,15} \approx 256$.
\item $\mathcal{D}(s) \Rightarrow s \in {256, \dots}$; $64,128,224 \notin \mathcal{S}$ (all fail $\mathcal{M}$ or record generation).
\item $\mathcal{C}(32768,256,n) \Rightarrow n \leq \lfloor 32768/256 \rfloor = 128$.
\item $\mathcal{F}: (\log N=15,s \geq 256,n \leq 128)$.
\end{itemize}

\begin{figure}[!htbp]
  \centering
  \includegraphics[width=\columnwidth]{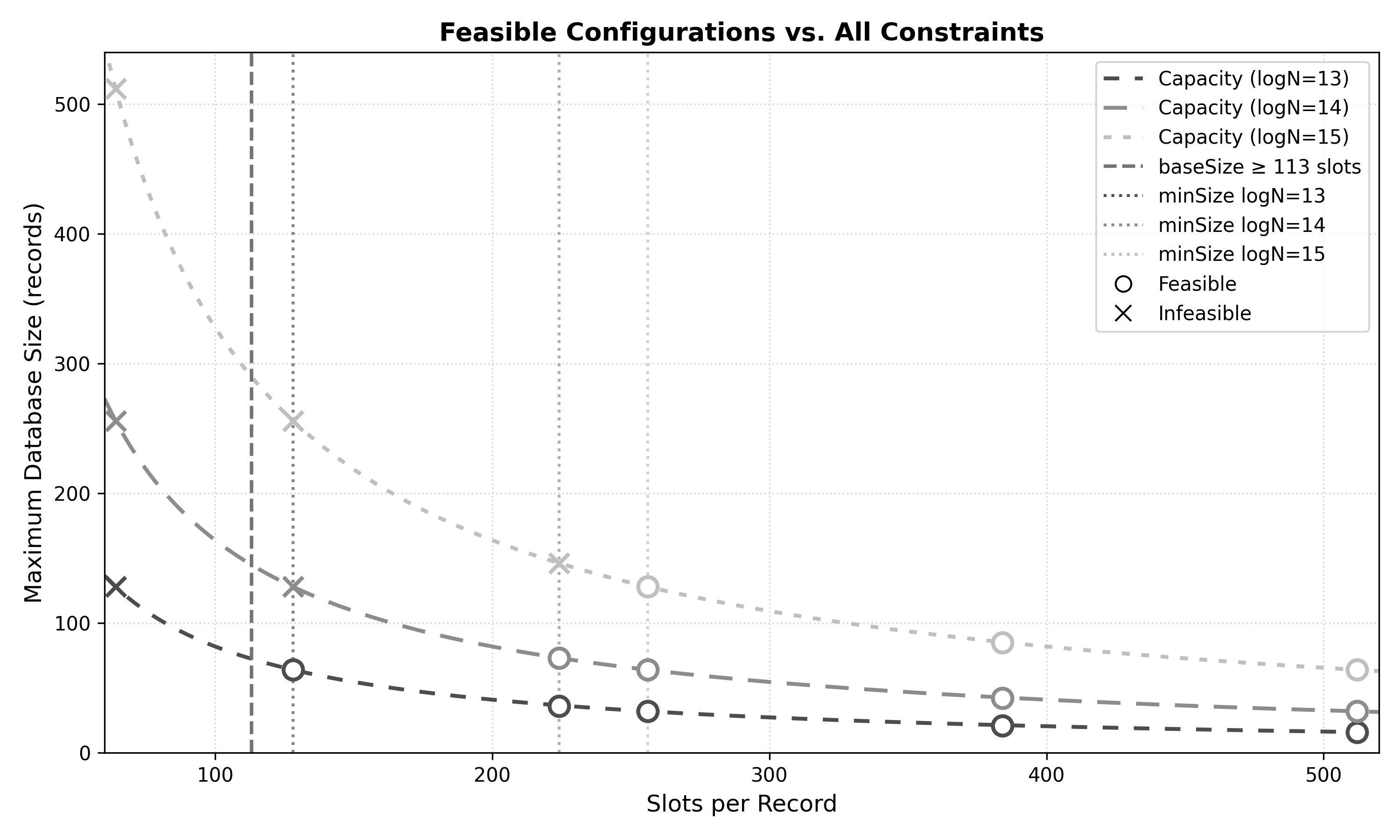}
  \caption{Feasible configurations under the joint constraints $\mathcal{C}$, $\mathcal{M}$, and $\mathcal{D}$. Dashed 
  curves show ring-capacity limits for $\log N \in \{13,14,15\}$, vertical lines mark template-driven minima $record_{\mu,\log N}$, 
  and x-axis ticks correspond to discrete slot sizes $\mathcal{S}$. Circles indicate feasible triples $(\log N,record_s,n)$.}
  \label{fig:exp-matrix}
\end{figure}

\noindent\textbf{Implications.} Increasing $\log N$ raises ring capacity $N$ and thus $n$, but also requires larger $record_s$ 
if given richer templates. Feasible configurations occur only where all 3 predicates are met, as depicted in Fig.~\ref{fig:exp-matrix}

\subsection{MULTI-CHANNEL ARCHITECTURE}
\begin{figure*}[!htbp]
  \centering
  \includegraphics[width=\textwidth]{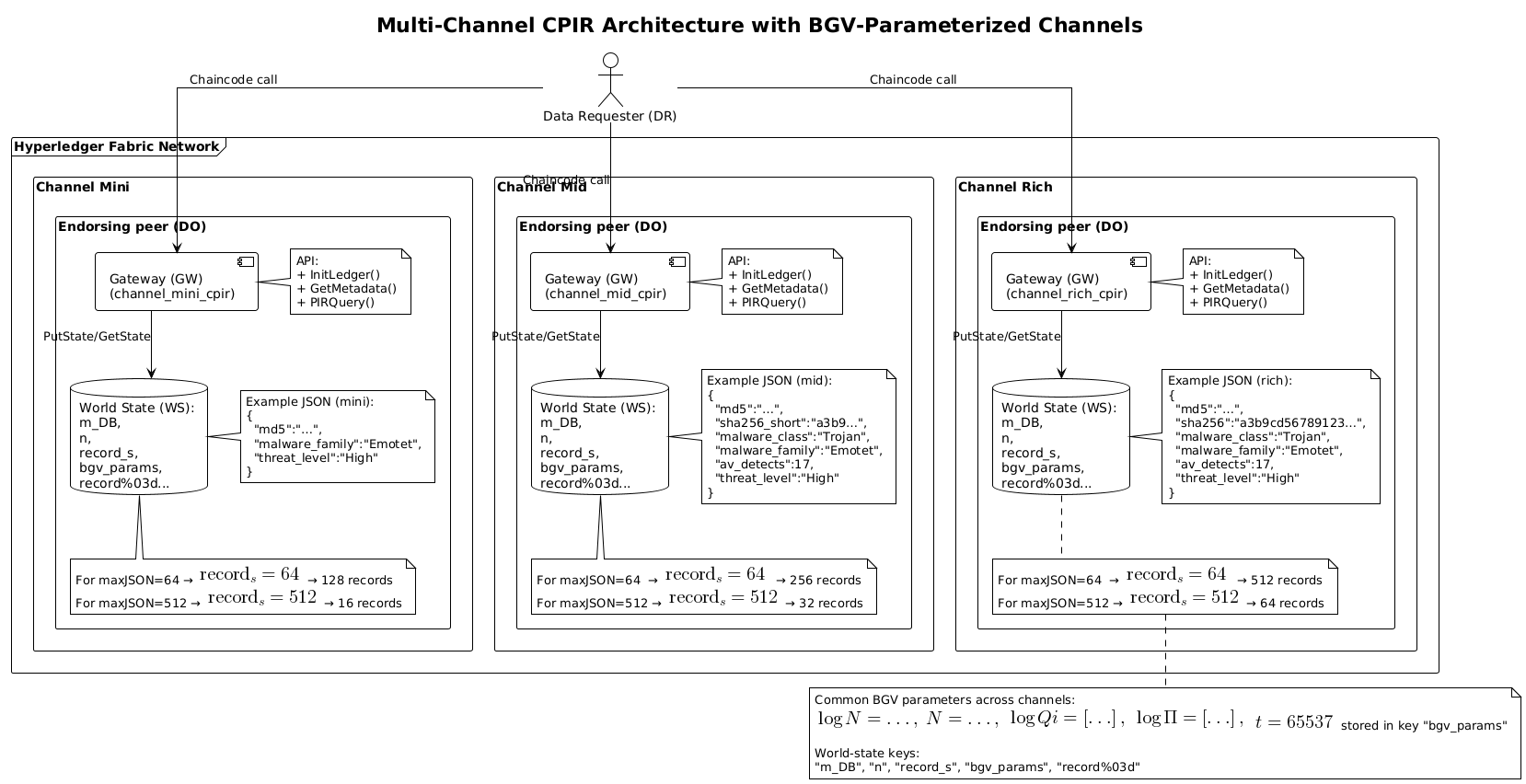}
  \caption{Multi-channel CPIR architecture. 
  Each channel instantiates a separate CPIR chaincode and maintains its own \emph{$m_{DB}$} polynomial, 
  parameterized by $\log N$. This allows compact, mid-size, and rich CTI records to coexist under the same Fabric network.}
  \label{fig:multi-channel}
\end{figure*}

The packing strategy and feasibility constraints highlight an important observation: 
no single homomorphic parameter set can efficiently support the full diversity of dynamic CTI records encountered in practice.
Compact records fit comfortably under smaller rings, while full JSON objects with long cryptographic hashes exceed the slot budget of these configurations. 
To balance scalability and expressiveness, we design a \emph{multi-channel architecture} in Hyperledger Fabric 
(Fig.~\ref{fig:multi-channel}), 
where each channel is provisioned with a distinct BGV parameter set and record template.

\textbf{a) Channel Mini} ($N=2^{13}$).  
Supports compact records with maximum scalability and lowest query latency. 
For example, with $N=8192$ slots, the system accommodates up to 128 records when $\max_i |d_i| \leq 64$ bytes, 
and 16 records when $\max_i |d_i| \leq 512$ bytes.

\textbf{b) Channel Mid} ($N=2^{14}$).
Targets medium-sized records that include MD5 and truncated SHA-256 fields alongside classification metadata. 
With $N=16384$ slots, the system supports up to 256 records at $\leq 64$ bytes or 32 records at $\leq 512$ bytes.

\textbf{c) Channel Rich} ($N=2^{15}$).
Handles the most detailed records, including full-length hashes and multiple metadata fields. 
Here, $N=32768$ slots allow up to 512 records at $\leq 64$ bytes or 64 records at $\leq 512$ bytes.

\noindent\textbf{Channel semantics.}  
As shown in Fig.~\ref{fig:multi-channel}, each channel maintains its own PIR chaincode instance and world state. 
The world state contains:
\begin{itemize}
    \item The \emph{polynomial view}: the packed plaintext polynomial \emph{$m_{DB}$} under key \emph{``m\_DB"}.
    \item The \emph{normal view}: JSON records stored optionally under keys \emph{``record\%03d"} for auditability.
    \item \emph{Metadata}: 
    \begin{itemize}
        \item \emph{``n"}: number of records $n$,
        \item \emph{``record\_s"}: slots per record $record_s$,
        \item \emph{``bgv\_params"}: $\{\log N, N, \log Q_i, \log P_i, T\}$.
    \end{itemize}
\end{itemize}

\subsection{WORKFLOW DETAILS}

The proposed system has five main routines (Alg.~\ref{alg:initledger}–\ref{alg:decryptresult}), corresponding to 3 chaincode-side functions 
(\texttt{InitLedger}, \texttt{GetMetadata}, \texttt{PIRQuery}) and 2 client-side utilities (\texttt{FormSelectionVector}, \texttt{DecryptResult}). 
Figure~\ref{fig:system-overview} provides the high-level overview. 
A Data Writer ($\mathcal{DW}$) provisions the database $D=\{d_0,\dots,d_{n-1}\}$, 
a Data Owner ($\mathcal{DO}$) maintains the polynomial \emph{$m_{DB}$} in world state and executes PIR evaluations, 
and a Data Requester ($\mathcal{DR}$) retrieves $d_i$ privately using homomorphic encryption.
The detailed steps are as follows:
\begin{enumerate}
    \item \textbf{$\mathcal{DW}$ submits initialization.}  
    $\mathcal{DW}$ calls \texttt{InitLedger} (Alg.~\ref{alg:initledger}) via $\mathcal{GW}$ (\emph{submit}) 
    with inputs $(n,\,record_{s}^{\mathcal{DW}})$ and an optional hint $(\log N,\log Q_i,\log P_i,T)$. 
    Here $record_{s}^{\mathcal{DW}}$ denotes the maximum JSON size anticipated by the writer. 

    \begin{algorithm}[!htbp]
      \caption{\texttt{InitLedger} (chaincode)}
      \label{alg:initledger}
      \begin{algorithmic}[1]
      \REQUIRE $n$; $record_s^{\mathcal{DW}}$; op: $\mathrm{hint}$
      \STATE $\log N \leftarrow \selMinLogN{\!\left(n,\,8 \cdot \left\lceil \tfrac{record_s^{\mathcal{DW}}}{8} \right\rceil\right)}$
      \IF {$\log N=\varnothing$} 
          \STATE \textbf{return} $\bot$ 
      \ENDIF
      \STATE $bgvParams \gets \selParams{(\log N,\, op:\mathrm{hint})}$
      \STATE $D \gets \genRecords{(n,\,record_s^{\mathcal{DW}})}$
      \IF {$\exists i:\ |d_i|>record_s^{\mathcal{DW}}$} 
          \STATE \textbf{return} $\bot$ 
      \ENDIF
      \STATE $record_s \gets 8 \cdot \left\lceil \dfrac{\max_i |d_i|}{8} \right\rceil$
      \IF{$\neg \feasible{\log N,n,record_s}$} 
          \STATE \textbf{return} $\bot$ // infeasible configuration
      \ENDIF

      \STATE $c \gets [0,\dots,0] \in \mathbb{Z}_T^N$ \hfill // init coefficient vector
      \FOR{$i \in [0,n-1]$}
          \STATE $J_i \gets \{i\cdot record_s,\dots,(i+1)\cdot record_s-1\}$ \hfill // slot window for $d_i$
          \FOR{$k=0$ \TO $record_s-1$}
              \IF{$k < |d_i|$}
                  \STATE $c[J_i[k]] \gets \textsf{byte}(d_i[k])$ \hfill // copy byte of record
              \ELSE
                  \STATE $c[J_i[k]] \gets 0$ \hfill // padding
              \ENDIF
          \ENDFOR
      \ENDFOR

      \STATE $\emph{$m_{DB}$}(X) \gets \encP{c} \in \mathbb{Z}_T[X]/(X^N+1)$
      \STATE $worldState \gets \{\emph{$m_{DB}$},\;n,\;record_s,\;bgvParams,\;op:D\}$
      \STATE \textbf{return} \textsc{ok}
      \end{algorithmic}
    \end{algorithm}

    \item \textbf{$\mathcal{DO}$ validates and derives parameters.}  
    $\mathcal{DO}$ rounds the writer’s input to a discrete slot size 
    $record_{s}^{\mathcal{GW}} = 8 \cdot \lceil record_{s}^{\mathcal{DW}} / 8 \rceil$. 
    If $\log N$ is absent, the smallest feasible $\log N$ is chosen such that 
    $\mathcal{C}(N,record_{s}^{\mathcal{GW}},n)$ holds. 
    BGV parameters $\{\log N,N,\log Q_i,\log P_i,T\}$ are constructed and stored.

    \begin{algorithm}[!htbp]
      \caption{\texttt{GetMetadata} (chaincode)}
      \label{alg:getmetadata}
      \begin{algorithmic}[1]
      \REQUIRE $\varnothing$
      \STATE $n \gets worldState.n$
      \STATE $record_s \gets worldState.record_s$
      \STATE $paramsMeta \gets worldState.bgvParams$
      \IF{$n=\varnothing \;\lor\; record_s=\varnothing \;\lor\; paramsMeta=\varnothing$}
      \STATE \textbf{return} $\bot$
      \ENDIF
  
      \STATE $paramsMeta=(\log N, N, \log Q_i[], \log P_i[], T)$
      \STATE $metadata \gets (n, record_s, paramsMeta)$
      \STATE \textbf{return} $metadata$
      \end{algorithmic}
    \end{algorithm}

    \item \textbf{$\mathcal{DO}$ prepares records.}  
    Records $D=\{d_0,\dots,d_{n-1}\}$ are ingested or synthesized with $|d_i|\le record_{s}^{\mathcal{DW}}$. 
    The definitive slot allocation is then fixed as 
    $record_s = 8 \cdot \lceil \max_i |d_i| / 8 \rceil$ (discrete policy), 
    checked against feasibility predicates 
    $\mathcal{M}(\log N,record_s)$, $\mathcal{D}(record_s)$, $\mathcal{C}(N,record_s,n)$. 

    \item \textbf{Pack and persist.}  
    Each $d_i$ is placed in a disjoint window $J_i=\{i\cdot record_s,\dots,(i+1)record_s-1\}$ of $c=(c_0,\dots,c_{N-1})$, 
    zeros pad unused slots, and the polynomial $\emph{$m_{DB}$}(X)=\sum c_j X^j$ is encoded at max level. 
    World state stores \emph{``m\_DB"}, \emph{``n"}, \emph{``record\_s"}, and \emph{``bgv\_params"} 
    plus optional \emph{``record\%03d"} entries. 

    \item \textbf{$\mathcal{DR}$ discovers metadata.}  
    $\mathcal{DR}$ calls \texttt{GetMetadata} (Alg.~\ref{alg:getmetadata}) via \emph{evaluate} 
    to obtain $(n,record_s,\log N,N,T,\log Q_i,\log P_i)$. 
    This enables reconstruction of the cryptographic context. 

    \item \textbf{$\mathcal{DR}$ instantiates crypto context.}  
    From metadata, $\mathcal{DR}$ builds parameters, executes $\emph{KeyGen}(\lambda)\!\to\!(pk,sk)$, 
    and prepares encoder/encryptor objects. 

    \begin{algorithm}[!htbp]
      \caption{\texttt{FormSelectionVector} (client)}
      \label{alg:encryptquery}
      \begin{algorithmic}[1]
      \REQUIRE $pk$; $i\in[n]$; $record_s$; $N$
      \IF{$i<0 \;\lor\; i\ge n$} \STATE \textbf{return} $\bot$ \ENDIF
      \IF{$n\cdot record_s > N$} \STATE \textbf{return} $\bot$ \ENDIF
      \STATE $J_i \gets \{i\cdot record_s,\dots,(i+1)\cdot record_s-1\}$
      \STATE ${v}_i \in \{0,1\}^{N} \gets \mathbf{0}$
      \FOR{$j \in J_i$} \STATE ${v}_i[j] \gets 1$ \ENDFOR \hfill // windowed selector
      \STATE $m_q(X) \gets \encP{{v}_i}$ \hfill // polynomial encode at max level
      \STATE $ct_q \gets \emph{Enc}_{pk}(m_q)$
      \STATE $ct_q^{B64} \gets \encB{\ser{ct_q}}$
      \STATE \textbf{return } $ct_q^{B64}$ \hfill // Base64(marshalled ciphertext)
      \end{algorithmic}
    \end{algorithm}

    \item \textbf{Form and encrypt query.}  
    For index $i\in[n]$, $\mathcal{DR}$ runs \texttt{FormSelectionVector} (Alg.~\ref{alg:encryptquery}): 
    define $J_i$, set ${v}_i$ with ones on $J_i$, encode to $m_q(X)$, encrypt as $ct_q=\emph{Enc}_{pk}({v}_i)$, 
    and serialize/Base64-encode. 

    \item \textbf{PIR query evaluation.}  
    $\mathcal{DR}$ issues \texttt{PIRQuery}($ct_q^{B64}$) (Alg.~\ref{alg:pirquery}) via \emph{evaluate}. 
    $\mathcal{DO}$ decodes, reloads \emph{$m_{DB}$} if necessary, and computes 
    $ct_r=\emph{Eval}(ct_q,\emph{$m_{DB}$})$, returning the Base64-encoded ciphertext.   

    \begin{algorithm}[!htbp]
      \caption{\texttt{PIRQuery} (chaincode)}
      \label{alg:pirquery}
      \begin{algorithmic}[1]
      \REQUIRE $ct_q^{B64}$
      \IF{$ct_q^{B64} = \varnothing$} 
          \STATE \textbf{return} $\bot$ 
      \ENDIF
      \STATE $ct_q \gets \des{\decB{(ct_q^{B64})}}$
      \IF{\emph{$m_{DB}$} not cached in memory}
          \STATE $\emph{$m_{DB}$} \gets worldState.\emph{$m_{DB}$}$
      \ENDIF
      \STATE $ct_r \gets \emph{Eval}(ct_q, \emph{$m_{DB}$})$
      \STATE $ct_r^{B64} \gets \encB{\ser{(ct_r)}}$
      \STATE \textbf{return} $ct_r^{B64}$
      \end{algorithmic}
    \end{algorithm}

    \item \textbf{Decryption and reconstruction.}  
    $\mathcal{DR}$ runs \texttt{DecryptResult} (Alg.~\ref{alg:decryptresult}) to recover $m'(X)$, 
    extract bytes from $J_i$, stop at padding zero, and reconstruct $d_i$.

    \begin{algorithm}[!htbp]
      \caption{\texttt{DecryptResult} (client)}
      \label{alg:decryptresult}
      \begin{algorithmic}[1]
      \REQUIRE $ct_r^{B64}$; $sk$; $i\in[n]$; $record_s$; $n$
      \STATE \textbf{if } $i<0$ \textbf{ or } $i\ge n$ \textbf{ then return } $\bot$
      \STATE \textbf{if } $n\cdot record_s > N$ \textbf{ then return } $\bot$ \hfill // sanity
      \STATE $ct_r \leftarrow \des{\decB{(ct_r^{B64})}}$
      \STATE $u \in \mathbb{Z}_T^{N} \leftarrow \decP{\,m'(X)\,} \leftarrow \emph{Dec}_{sk}(ct_r)$
      \STATE $J_i \leftarrow \{i\cdot record_s,\dots,(i+1)\cdot record_s-1\}$
      \STATE $b \leftarrow$ byte array // init empty buffer for record
      \FOR{$j \in J_i$}
          \IF{$u[j]=0$} \STATE \textbf{break} \ENDIF \hfill // stop at padding zero
          \STATE $b.\mathsf{append}(u[j])$ 
      \ENDFOR
      \STATE \textbf{if } {$record_s=1$} \textbf{ then return } $u[i\cdot record_s]$
      \STATE $d_i \leftarrow {dec^{UTF8}}(b)$ 
      \STATE \textbf{return } $d_i$
      \end{algorithmic}
  \end{algorithm}
\end{enumerate}

\noindent\textit{Remark (Single-record PIR).}
The proposed system currently supports windowed selection vectors for single-record PIR queries.
Extending to multi-record queries is feasible by modifying the selection vector
to include multiple windows, though this increases ciphertext size and evaluation complexity.
For simplicity, we focus on single-record retrieval in this work.

\section{PERFORMANCE EVALUATION}\label{sec:evaluation}
\subsection{EXPERIMENTAL SETUP}

All experiments were executed on a local Ubuntu~24.04 host running under WSL2 on an Intel~Core~i5-3380M~CPU 
(2~cores/4~threads, 2.90~GHz) with 7.7~GB of RAM and a 1~TB~SSD. 
During evaluation, the average available memory was 6.9~GB with a 2~GB swap partition, 
and the root filesystem reported 946~GB of free space.
The software stack consisted of Go~1.24.1, Docker~27.4.0, and Docker~Compose~v2.31.0, 
hosting Hyperledger~Fabric~v2.5 with LevelDB as the world-state database and a Solo ordering service, using recommended settings 
(\emph{BatchSize.AbsoluteMaxBytes}=99~MB, \emph{PreferredMaxBytes}=2~MB per block).
The network configuration comprised a single organization with one peer per channel, 
sufficient for privacy evaluation, though it can be extended to multiple peers and organizations as needed.

Our implementation employs Fabric~GO~SDK~\cite{fabric_gateway_pkg_client_nodate} for blockchain interactions and
the \emph{Lattigo}~v6 library~\cite{noauthor_lattigo_2024} as the homomorphic encryption backend.
Lattigo provides a Go-native implementation of the BGV scheme~\cite{cryptoeprint:2011/277},
whose security and correctness have been validated in prior literature. 
Accordingly, our focus is on evaluating its \emph{practical performance within a permissioned blockchain environment} for 
enabling \emph{private reads} from world state, rather than re-verifying the theoretical properties of BGV itself.

Unless otherwise stated, each reported value represents the mean of 20~executions under both cold and warm cache conditions, 
cross-verified against peer logs for consistency.

\subsection{CRYPTOGRAPHIC PERFORMANCE}

\noindent\textbf{Parameter Configuration.}
Table~\ref{tab:params} summarizes the BGV parameter sets used in our evaluation, 
including the ring dimension $N$, ciphertext modulus chain $(\log Q_i, \log P_i)$, 
plaintext modulus $T$, slot allocation $record_s$, and database size $n$. 
Each configuration corresponds to one Fabric channel and maintains a feasible 
packing ratio as defined in Section~\ref{sec:proposed_system}.

\begin{table}[!htbp]
\caption{Default BGV Parameter Configuration per Channel}
\label{tab:params}
\centering
\begin{tabular}{|c|c|c|c|c|c|}
\hline
$N$ & $\log Q_i$ & $\log P_i$ & $T$ & $record_s$ & $n$  \\
\hline
$2^{13}$ & [54] & [54] & 65537 & 128 & 64 \\
$2^{14}$ & [54] & [54] & 65537 & 224 & 73 \\
$2^{15}$ & [54] & [54] & 65537 & 256 & 128 \\
\hline
\end{tabular}
\end{table}

\noindent\textbf{Overall Results.}
Table~\ref{tab:crypto-times} lists the measured execution time of core cryptographic operations 
(\emph{KeyGen}, \emph{Enc}, \emph{Eval}, and \emph{Dec}) for each evaluated ring dimension~$N$. 
\begin{table}[!htbp]
  \caption{Execution Time of Cryptographic Operations (ms)}
  \label{tab:crypto-times}
  \centering
  \begin{tabular}{|c|c|c|c|c|}
  \hline
  $N$ & \emph{KeyGen} & \emph{Enc} & \emph{Eval} & \emph{Dec} \\
  \hline
  $2^{13}$ & 27.7 & 11.7 & 16.9 & 5.1 \\
  $2^{14}$ & 42.7 & 27.7 & 30.7 & 11.9 \\
  $2^{15}$ & 55.0 & 49.8 & 64.8 & 15.6 \\
  \hline
  \end{tabular}
\end{table}

Table~\ref{tab:artifact-sizes} summarizes the corresponding sizes of primary cryptographic artifacts, 
including public and secret keys, ciphertexts ($ct_q$, $ct_r$), the encoded plaintext database 
$m_{\mathrm{DB}}$, and auxiliary metadata. 
\begin{table}[!htbp]
  \caption{Size of Main Cryptographic Artifacts (KB)}
  \label{tab:artifact-sizes}
  \centering
  \begin{tabular}{|c|c|c|c|c|c|c|}
  \hline
  $N$ & pk & sk & $ct_q$ & $ct_r$ & \emph{$m_{DB}$} & Metadata \\
  \hline
  $2^{13}$ & 256.1 & 128.0 & 128.3 & 128.3 & 64.3 & 0.08 \\
  $2^{14}$ & 512.1 & 256.0 & 256.3 & 256.3 & 128.3 & 0.08 \\
  $2^{15}$ & 1024.1 & 512.0 & 512.3 & 512.3 & 256.3 & 0.08 \\
  \hline
  \end{tabular}
\end{table}

\begin{figure}[!htbp]
\centering
\includegraphics[width=\columnwidth]{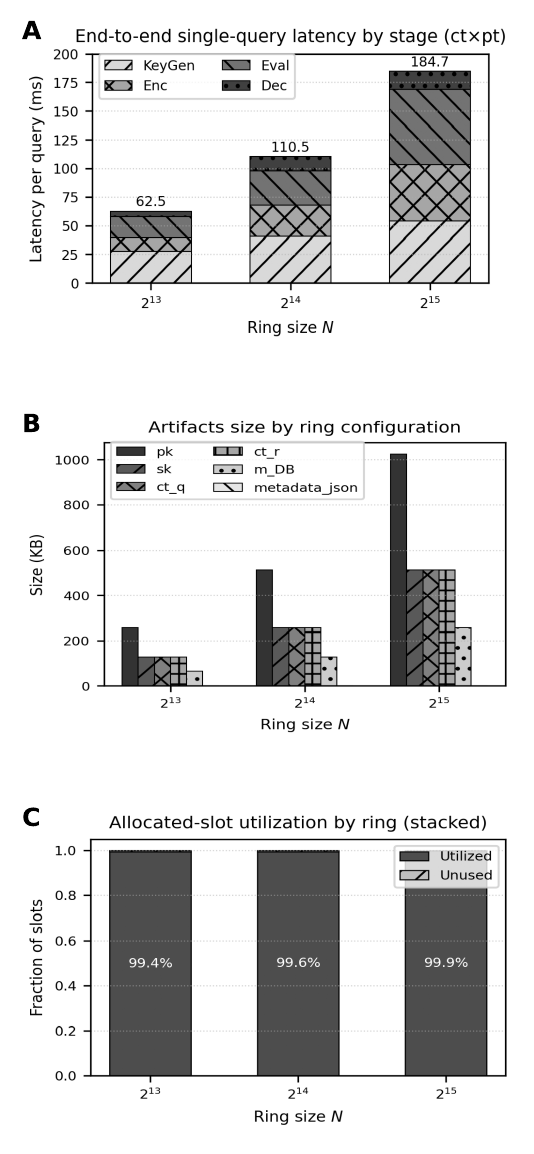}
\caption{Cryptographic performance of the BGV-based CPIR system: 
(A)~latency by algorithmic stage, (B)~artifact size by ring configuration, and 
(C)~allocated-slot utilization.}
\label{fig:latency-stages}
\end{figure}

Figure~\ref{fig:latency-stages} consolidates the main cryptographic evaluation metrics:
(A)~end-to-end latency by algorithmic stage,
(B)~serialized artifact size, and
(C)~slot utilization ratio within the packed database polynomial \emph{$m_{DB}$}.

\subsection{BLOCKCHAIN PERFORMANCE}

\begin{figure}[!htbp]
  \centering 
  \includegraphics[width=\columnwidth]{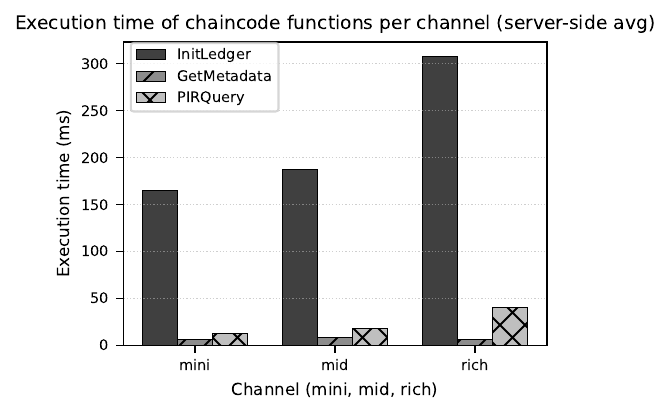} 
  \caption{Average chaincode execution time by function and ring size. Each bar represents the 
  mean execution time over multiple epochs.} \label{fig:chaincode-timings} 
\end{figure}

\noindent\textbf{Chaincode Execution Timings.} Figure~\ref{fig:chaincode-timings} and Table~\ref{tab:chaincode-timings} summarize the 
average server-side execution time of the main chaincode functions across three channels, each corresponding to a different ring size $N$:

\begin{table}[!htbp]
  \caption{Average Chaincode Execution Time (ms)}
  \label{tab:chaincode-timings}
  \centering
  \begin{tabular}{|c|c|c|c|}
  \hline
  $N$ & \texttt{InitLedger} & \texttt{GetMetadata} & \texttt{PIRQuery} \\
  \hline
  $2^{13}$ & 165.24 & 6.16 & 12.53 \\
  $2^{14}$ & 187.03 & 8.31 & 18.17 \\
  $2^{15}$ & 307.87 & 5.94 & 40.36 \\
  \hline
  \end{tabular}
\end{table}

\noindent\textit{Remark (execution paths).}  
Transactions that modify world state (e.g., \texttt{InitLedger}) are issued via the 
\emph{submit} path and committed through Solo ordering,
while read-only operations (\texttt{GetMetadata}, \texttt{PIRQuery}) use the 
\emph{evaluate} path, bypassing block creation and ordering.  

\begin{figure*}[!ht]
  \centering
  \includegraphics[width=\textwidth]{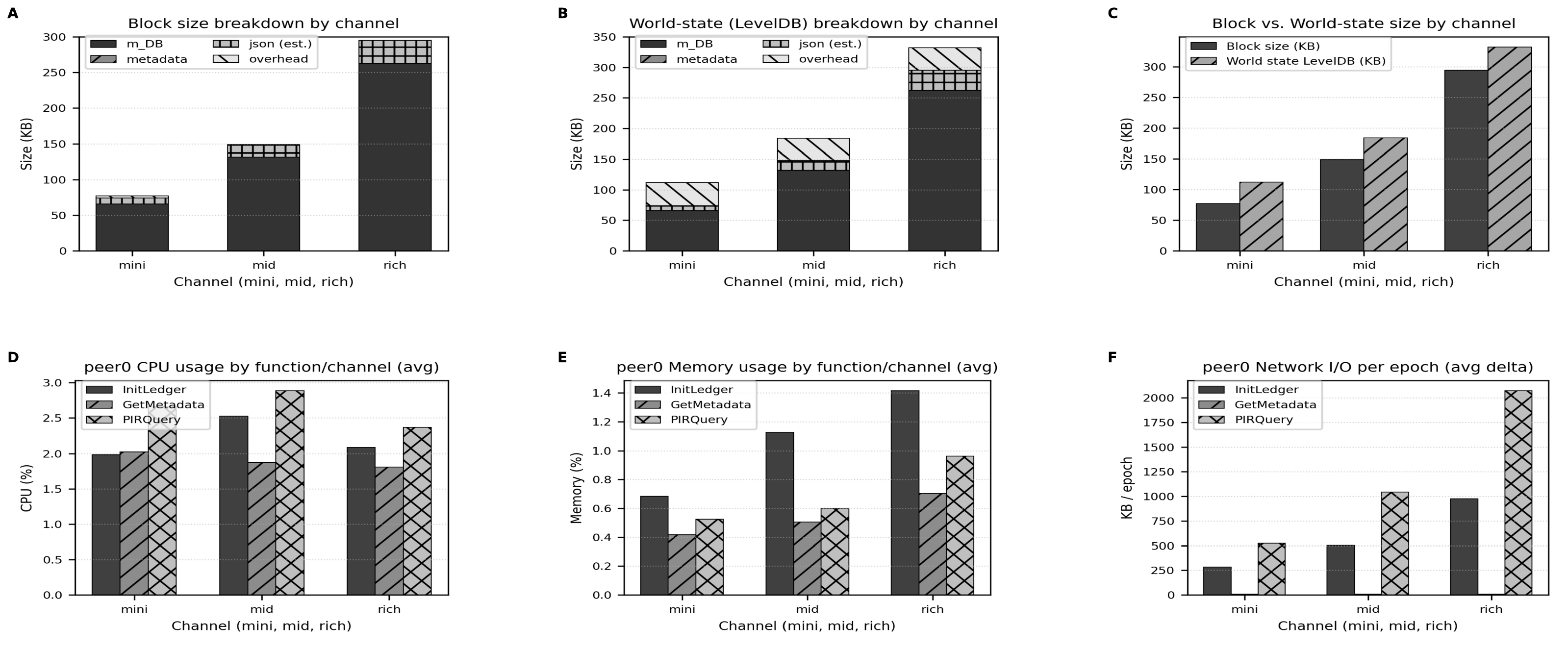}
  \caption{Blockchain-side evaluation of the CPIR system across three Fabric channels.
  (A–C)~Storage-level metrics: block and world-state composition. 
  (D–F)~Peer-level resource utilization: CPU, memory, and network I/O per function.}
  \label{fig:blockchain-metrics}
\end{figure*}

\noindent\textbf{Blockchain performance.} Figure~\ref{fig:blockchain-metrics} summarizes the performance of the 
CPIR chaincode within Hyperledger~Fabric across three channels, each corresponding to a different ring size $N$ and 
record configuration: 
(A)~block size breakdown,
(B)~world-state (LevelDB) breakdown,
(C)~block vs. world-state size,
(D)~peer CPU utilization,
(E)~peer memory utilization, and
(F)~peer network I/O.
On-chain network behavior is reported in Table~\ref{tab:pirquery-netio}, which presents the average peer-side 
network I/O per \texttt{PIRQuery} transaction across the three channel configurations.

\begin{table}[!htbp]
  \centering
  \caption{Peer network I/O per PIRQuery (avg, KB/tx)}
  \label{tab:pirquery-netio}
  \begin{tabular}{|c|c|c|}
  \hline
  Channel & $N$ & NET I/O (KB/tx) \\
  \hline
  mini  & $2^{13}$ & 524.62 \\
  mid   & $2^{14}$ & 1042.09 \\
  rich  & $2^{15}$ & 2072.88 \\
  \hline
  \end{tabular}
\end{table}

\subsection{OVERALL SYSTEM PERFORMANCE}

Block and world-state sizes are detailed in Table~\ref{tab:block-state-size}, showing the breakdown of storage components per channel.

\begin{table*}[!ht]
  \centering
  \caption{Block and World-State Size per Channel Configuration}
  \label{tab:block-state-size}
  \begin{tabular}{|c|c|c|c|c|c|c|c|c|}
  \hline
  $N$ & \textbf{$n$} &
  $m_{\mathrm{DB}}$ (KB) & Metadata (KB) & JSON (KB) &
  Overhead$_{block}$ (KB) & Overhead$_{ws}$ (KB) &
  Block (KB) & World (KB) \\
  \hline
  $2^{13}$ & 64  & 65.838 & 0.061 & 8.064 & 3.037 & 38.037 & 77  & 112 \\
  $2^{14}$ & 73  & 131.374 & 0.062 & 16.206 & 1.358 & 36.358 & 149 & 184 \\
  $2^{15}$ & 128 & 262.446 & 0.063 & 32.512 & 0.000 & 36.979 & 294 & 332 \\
  \hline
  \end{tabular}
\end{table*}

\noindent\textbf{End-to-End Workflow Summary.}
Based on the data in Table \ref{tab:crypto-times} and Table \ref{tab:chaincode-timings}, 
we now to summarize two main operational workflows in the system, involving $\mathcal{DW}$ and $\mathcal{DR}$, 
while $\mathcal{DO}$ is implicitly involved as the executing peer during private queries.

The first workflow corresponds to $\mathcal{DW}$ initializing the ledger through \texttt{InitLedger}, 
which packs and encodes the plaintext database \emph{$m_{DB}$} into world state. 

The second workflow corresponds to $\mathcal{DR}$ performing a private query by sequentially 
executing \texttt{GetMetadata} → \emph{KeyGen} → \emph{Enc} → \texttt{PIRQuery} → \emph{Dec}.
The PIR evaluation itself is performed by $\mathcal{DO}$ (endorsing peer) using ciphertext-plaintext 
multiplication during the evaluate transaction.

Table~\ref{tab:end2end} summarizes the cryptographic and blockchain timings for both workflows, 
averaged across all three channel configurations.

\begin{table}[!htbp]
    \caption{End-to-End Performance Analysis (ms)}
    \label{tab:end2end}
    \centering
    \begin{tabular}{|c|c|c|c|}
    \hline
    {Workflow} & 
    \shortstack{{Cryptographic} \\ {Operations (ms)}} & 
    \shortstack{{Blockchain} \\ {Operations (ms)}} & 
    \shortstack{{Total Time} \\ {(ms)}} \\
    \hline
    $\mathcal{DW}$'s Upload  & 0.0 & 220.0 & 220.0 \\
    $\mathcal{DR}$'s Query  & 82.4 & 30.5 & 112.9 \\
    \hline
    \end{tabular}
\end{table}

\begin{figure}[!htbp] 
    \centering 
    \includegraphics[width=\columnwidth]{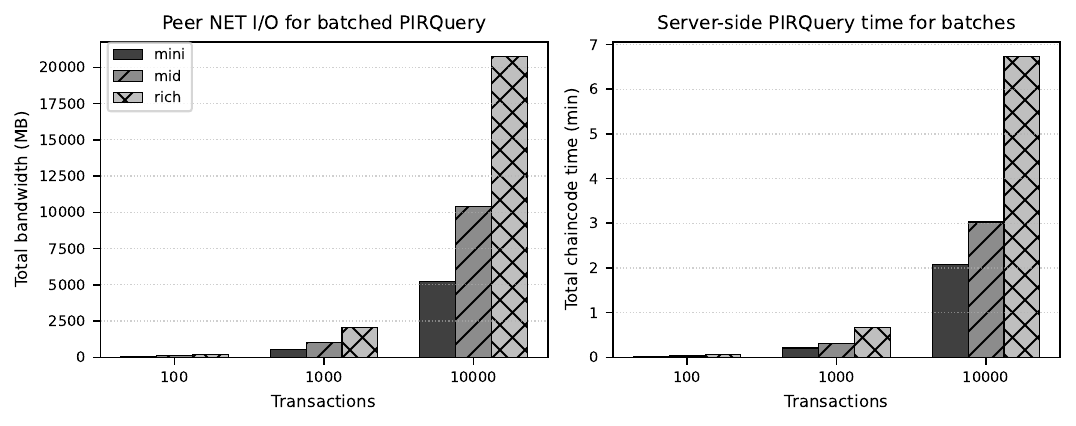} 
    \caption{Projected cost for multiple PIR queries at $2^{15}$ (peer perspective)} 
    \label{fig:batch-cost-rich} 
\end{figure}

\noindent\textbf{Projected Cost for PIR Query.}
Among our results, for the ring size $2^{15}$ configuration, issuing 100/1000/10000 private queries incurs 
$\approx $207MB / 2.07 GB / 20.7 GB of peer-side network I/O with an associated chaincode evaluation time 
of $\approx$0.07/0.67/6.73 minutes, respectively, when using our protocol, as shown in Table \ref{tab:batch-cost-rich} 
and Figure \ref{fig:batch-cost-rich}.

\begin{table}[!htbp]
    \centering
    \caption{Projected cost for multiple PIR queries at $2^{15}$ (peer perspective)}
    \label{tab:batch-cost-rich}
    \begin{tabular}{|c|c|c|}
    \hline
    \# Transactions & Total Bandwidth (MB) & Total Time (min) \\
    \hline
    100   & 207.3  & 0.07 \\
    1000  & 2073.0 & 0.67 \\
    10000 & 20730.0 & 6.73 \\
    \hline
    \end{tabular}
\end{table}

\noindent\textit{Remark (on network I/O measurement).} 
The network I/O values in Table~\ref{tab:pirquery-netio} and Table~\ref{tab:batch-cost-rich} were obtained from \emph{docker stats} 
snapshots collected for the \emph{peer0.org1.example.com} and \emph{orderer0.group1.orderer.example.com} containers during each benchmark epoch. 

\begin{table*}[!ht]
  \centering
  \caption{Comparison with existing privacy-preserving query systems}
  \label{tab:comparison-related}
      \begin{tabular}{|c|c|c|c|c|c|c|c|c|}
      \hline
      \textbf{Work} &  
      \shortstack{\textbf{Targeted} \\ \textbf{System}} &
      \shortstack{\textbf{Privacy} \\ \textbf{Technique}} &
      \shortstack{\textbf{Execution} \\ \textbf{Location}} &
      \shortstack{\textbf{Targeted} \\ \textbf{Operation}} &
      \shortstack{\textbf{Comm. Cost} / \\ \textbf{Query (MB)}} &
      \shortstack{\textbf{E2E Query} \\ \textbf{Time (ms)}} &
      \shortstack{\textbf{Db Size} / \\ \textbf{(\# of records)}} &
      \shortstack{\textbf{Record Size} / \\ \textbf{length (bytes)}} \\
      \hline
      \cite{kumar_debpir_2025} & HLF  & OT & on-chain & read / write & N/A & 600 & 10,000 & N/A \\
      \cite{kumar_bron_2024} & HLF & ZKP & on-chain & read / write & N/A & N/A & 10,000 & N/A \\
      \cite{xiao_cloak_2024} & Blockchain  & DPF & off-chain & read & 0.0002-0.0005 & 0.4-0.5 & 4-64 & 16-32 \\
      \cite{mazmudar_peer2pir_2025} & IPFS & HE-PIR & off-chain & read & 10-15 & 500-16,000 & 1,000-1M & 256,000 \\
      \cite{Kaihua2019applying} & Bitcoin & Hybrid PIR & off-chain & read & 0.65 & 2840 & 7,688-512,460 & 62-876 \\
      \textbf{Ours} & \textbf{HLF} & \textbf{HE-PIR} & \textbf{on-chain} & \textbf{read} & \textbf{1-2} & \textbf{112.9} & \textbf{16-512} & \textbf{64-512} \\
      \hline
      \end{tabular}
\end{table*}

\begin{figure*}[!htbp]
  \centering
  \includegraphics[width=\textwidth]{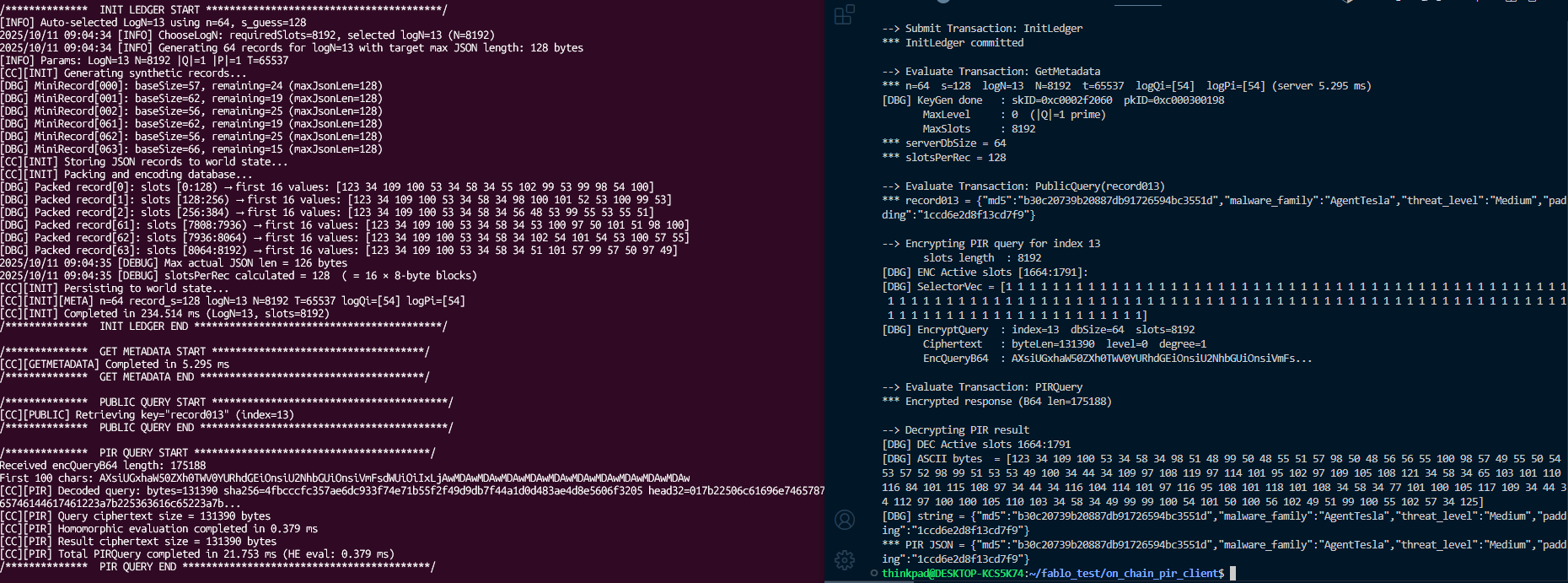}
  \caption{Detailed execution logs illustrating the privacy-preserving query workflow.
  Left: peer-side chaincode logs showing InitLedger, GetMetadata, and PIR evaluation with 
  ciphertext-plaintext multiplication.
  Right: client-side logs showing metadata retrieval, query encryption, PIR query invocation, 
  and decryption of the result.}
  \label{fig:query-privacy}
\end{figure*}

\section{DISCUSSION}\label{sec:discussion}
\subsection{FUTURE WORK AND OUTLOOK}
Figure \ref{fig:query-privacy} presents detailed execution logs from both the peer (left) and client (right) sides,
demonstrating the feasibility of bringing \emph{private reads} to Hyperledger Fabric via HE-based PIR.
Although our prototype validates this concept, several challenges remain in performance, scalability, and deployment.
Future research should address \emph{(i)} side-channel resilience through constant-time chaincode evaluation and standardized 
ciphertext serialization to mitigate timing and size-based leakages. 
\emph{(ii)} Scaling beyond moderate database sizes will require some form of sublinear or sharded CPIR, 
similar to~\cite{simplepir_2023,piano_pir_2024,hintless_pir_2024,ypir_2024}, yet adapted to Fabric’s architecture and resource constraints.
Moreover, \emph{(iii)} extending the current ciphertext–plaintext (\emph{ct}×\emph{pt}) evaluation to fully encrypted-database (\emph{ct}×\emph{ct}) 
computation would enable richer on-chain analytics under encryption, though it demands relinearization, rotation, and modulus-switching 
capabilities within Fabric’s resource limits. 
\emph{(iv)} Dynamic ledger updates, (\texttt{addRecord}) operation would allow incremental growth without full reinitialization,
raising questions around re-encoding and key management. 
\emph{(v)} Selective field-level retrieval through adaptive packing could enable varying length records and partial access control.
\emph{(vii)} Deployment optimization is crucial. In our setup, the chaincode ran in \emph{development mode} as an external service, 
causing duplicated gRPC transmissions for each ciphertext (\emph{ct\textsubscript{q}}, \emph{ct\textsubscript{r}}) due to two network 
hops—client$\leftrightarrow$peer and peer$\leftrightarrow$chaincode. This resulted in $\approx$2072~KB per query for $N=2^{15}$, 
as shown in Table~\ref{tab:batch-cost-rich}. Running the chaincode \emph{in-process} within the peer would eliminate the second hop, 
halving the network cost to $\approx$1024~KB per query.

\subsection{RELATED WORK COMPARISON}
Table~\ref{tab:comparison-related} contrasts our proposed PIR-based \emph{private reads} mechanism with several recent works 
discussed in Section~\ref{sec:introduction},
that integrate Private Information Retrieval (PIR) or related cryptographic mechanisms 
to achieve query privacy across different domains.
The comparison focuses on six key aspects:
\emph{(i)} which target system the privacy solution is built for,
\emph{(ii)} the underlying privacy primitive (CPIR, IT-PIR, hybrid, DPF etc),
\emph{(iii)} whether privacy computation occurs on-chain (via smart contracts or chaincode) or off-chain,
\emph{(iv)} the targeted operation (read-only or read/write),
\emph{(v)} communication cost per query (in MB),
\emph{(vi)} end-to-end query time (in ms),
\emph{(vii)} database size (number of records), and
\emph{(viii)} record size (in bytes).

\section{CONCLUSION}\label{sec:conclusion}
This paper presented a Private Information Retrieval (PIR) mechanism for enabling \emph{private reads} in Hyperledger~Fabric, 
allowing endorsing peers to evaluate encrypted queries without learning which record was accessed. 
The proposed chaincode performs ciphertext–plaintext (\emph{ct×pt}) homomorphic multiplication directly within 
\emph{evaluate} transactions, preserving Fabric’s endorsement and audit semantics. 
Our prototype achieves an average end-to-end latency of 113~ms and a peer-side execution time below 42~ms, 
with approximately 2~MB of peer network traffic per query in development mode—reducible to about 1~MB under in-process deployment. 
Storage profiling shows near-linear growth in both block and world-state sizes as the ring dimension scales, 
and parameter analysis confirms practical support for up to 512~records of 64~bytes each under $2^{15}$.  
These results validate the feasibility of PIR-based \emph{private reads} in permissioned ledgers, 
offering millisecond-scale query latency and full compatibility with Fabric’s architecture. 
Future work should explore constant-time and sublinear execution, sharded CPIR architectures, and fully encrypted 
(\emph{ct×ct}) on-chain evaluation to further enhance scalability and privacy.

\bibliographystyle{IEEEtran}
\bibliography{bibliography}

\end{document}